\documentclass[pss]{wiley2sp}
\usepackage{amsmath}

\begin{document}

\title{Exciton dephasing in quantum dots: Coupling to LO phonons via
excited states}

\titlerunning{Exciton dephasing in quantum dots}


\author{%
  E. A. Muljarov\textsuperscript{\textsf{\bfseries 1,\Ast}},
  R. Zimmermann\textsuperscript{\textsf{\bfseries 2}}}

\authorrunning{E. A. Muljarov and R. Zimmermann}
\mail{e-mail
  \textsf{muljarov@gpi.ru}}

\institute{%
  \textsuperscript{1}\,General Physics Institute, Russian Academy of Sciences,
Vavilova 38, Moscow 119991, Russia\\
  \textsuperscript{2}\,Institut f\"ur Physik der
Humboldt-Universit\"at zu Berlin, Newtonstra\ss e 15, 12489
Berlin, Germany}

\received{XXXX, revised XXXX, accepted XXXX} \published{XXXX}


\abstract{
 We have found a novel mechanism of spectral
broadening and dephasing in quantum dots (QDs) due to the coupling
to longitudinal-optical (LO) phonons. In theory, this mechanism
comes into play only if the complete manifold of exciton levels
(including those in the wetting-layer continuum) is taken into
account. We demonstrate this nontrivial dephasing in different
types of QDs, using the exactly solvable quadratic coupling model,
here generalized to an arbitrary number of excitonic states. }
\maketitle                   

While acoustic phonons are known to produce a spectral broadening
of the exciton zero phonon line (ZPL) in QD spectra, it is widely
believed that the coupling to bulk dispersionless LO phonons forms
everlasting mixed states (exciton polarons) showing no line
broadening~\cite{Hameau99,Stauber06}. This is indeed true if the
model is restricted to a limited number of exciton states.
However, including a sufficiently large number of states results
in spectral broadening and dephasing. We have recently
demonstrated~\cite{Muljarov07} this nontrivial and quite
unexpected mechanism of dephasing using the exactly solvable
quadratic coupling model earlier developed for acoustic
phonons~\cite{Muljarov04}. This new mechanism is in clear contrast
with the acoustic phonon-induced dephasing as the latter provides
a finite ZPL width due to the phonon-mediated coupling already to
the next exciton level.

In this contribution, we derive the quadratic coupling model for a
single level (exciton ground state) from the standard linear
coupling between exciton levels in a QD and bulk LO phonons.
Exciton excited states then contribute effectively through virtual
transitions. We have generalized the exact solution of the model
with two levels~\cite{Muljarov06} to an arbitrary number of
states~\cite{Muljarov07}. Here we show explicitly a
diagonalization of the quadratic Hamiltonian and then calculate
the dephasing in different types of QDs.

We derive the model of quadratic coupling to LO phonons from the
standard Hamiltonian with a linear exciton-phonon coupling:
\begin{eqnarray}
H&=&\omega_0\sum_{\bf q} \left( a^\dagger_{\bf q} a_{\bf q}
+\frac{1}{2}\right)+ \sum_n E_n |n\rangle \langle n| \nonumber\\
&& + \sum_{nm} |n\rangle \langle m| \sum_{\bf q} M_{nm}({\bf q})
(a_{\bf q}+a^\dagger_{-\bf q})\,,
 \label{H}
\end{eqnarray}
where $\omega_0$ is the dispersionless LO phonon frequency, $E_n$
is the bare transition energy of a single-exciton state
$|n\rangle$ in a QD, \mbox{$M_{nm}({\bf q})\propto q^{-1}\langle
n|e^{i {\bf qr}_e}-e^{i {\bf qr}_h}|m\rangle$} are matrix elements
of the linear exciton-phonon Fr\"ohlich interaction, and
$\hbar=1$. This problem has been solved exactly in case of a few
exciton levels~\cite{Stauber06}, but generally it is not solvable
for an arbitrary number of levels because of the level-nondiagonal
coupling which is, at the same time, of key importance for the
dephasing. To preserve this coupling, on the one hand, and to
bring the problem to an exactly solvable form, on the other, we
make the following unitary transformation
\begin{equation}
H' = e^S H e^{-S} = H + [S, H] + [S,[S,H]]/2+ \dots \,,
\end{equation} where
the transformation operator $S$ has the form:
\begin{eqnarray}
S&=&\sum_{n\neq m}  |n\rangle \langle m| \sum_{\bf q} M_{nm}({\bf
q}) \nonumber\\
&&\times\left(\frac{a_{\bf
q}}{E_n-E_m-\omega_0}+\frac{a^\dagger_{-\bf
q}}{E_n-E_m+\omega_0}\right)\,.
 \label{S}
\end{eqnarray}
This transformation eliminated nondiagonal terms in first
order~\cite{Muljarov04}, by mapping them into diagonal ones which
are however quadratic in phonon operators. Higher powers of
$M_{nm}({\bf q})$ are also generated in the transformed
Hamil\-to\-ni\-an. However, far from resonance where the quadratic
coupling model is valid, they lead only to small quantitative
corrections and thus can be safely neglected.

 For the ground exciton state ($n=1$) the resulting
quad\-ra\-tic Hamiltonian takes the form:
\begin{eqnarray}
 H_1&=& E'_1+\omega_0\sum_{\bf q} \left(
a^\dagger_{\bf q} a_{\bf q} +\frac{1}{2}\right) \nonumber\\
&& + \sum_{\bf q} M_{11}({\bf q})\,( a_{\bf q} + a_{\bf
-q}^\dagger) \nonumber\\
&&- \frac{1}{2} \sum_{\bf p\, q} Q({\bf p},{\bf q}) \,( a_{\bf p}
+ a_{\bf -p}^\dagger ) ( a_{\bf q} + a_{\bf -q}^\dagger )\,,
 \label{H1}
\end{eqnarray}
where $$E'_1=E_1+\sum_{n\neq1}\sum_{\bf q}\frac{|M_{1n}({\bf
q})|^2\omega_0}{[(E_n-E_1)^2-\omega_0^2]}$$ is a renormalized
(polaron shifted) energy of the exciton ground state, and
\begin{eqnarray}
Q({\bf p, q})&=& 2\sum_{n\neq1}
\frac{E_n-E_1}{(E_n-E_1)^2-\omega_0^2} \,M_{1n}({\bf p})
M_{n1}({\bf q})\nonumber\\
&\equiv&\sum_{n\neq1} F_n({\bf p}) F_n({\bf q}) \label{Q}
\end{eqnarray}
is the kernel of the quadratic coupling, here expressed in terms
of new functions $F_n({\bf q})$. Note that the Hamiltonian
Eq.(\ref{H1}) contains also the state-diagonal linear coupling
$M_{11}$.

The absorption spectrum is calculated as a Fourier transform of
the linear response on a delta-pulse excitation,
\begin{equation}
P(t)=\theta(t)\langle e^{iH_0t} e^{-iH_1t}\rangle,
\end{equation}
where a finite-temperature expectation value is taken over the
unperturbed phonon system
$$
H_0=\omega_0\sum_{\bf q} \left( a^\dagger_{\bf q} a_{\bf q} +
\frac{1}{2}\right).
$$
To do so we follow the method of Ref.~\cite{Stauber06} and
diagonalize $H_1$ directly in a few successive steps.

First of all, we make a shift transformation:
\begin{equation}
b_{\bf q}=a_{\bf q}+f_{\bf q}
\end{equation}
 ($f_{\bf q}$ will be given later).
This is only to eliminate the linear coupling $M_{11}$, all other
terms in Eq.(\ref{H1}) remaining the same, except for a new
renormalization of the transition energy to $\widetilde{E}_1=E'_1
+E_{\rm pol}^{(2)}$ with an additional polaron shift $E_{\rm
pol}^{(2)}=-\sum_{\bf q}M_{11}({\bf q})f_{\bf q}$. The next step
is a rotation of the phonon basis, in order to make the quadratic
coupling diagonal. To do this we diagonalize the Hermitian matrix
($E_n-E_1>\omega_0$ is assumed for simplicity):
\begin{equation}
A_{nm}\equiv\sum_{\bf q} F_n(-{\bf q}) F_m({\bf q})= \sum_\nu
\Lambda_\nu\, X_{\nu n} X^\ast_{\nu m}
 \label{A}
\end{equation}
and introduce new phonon operators:
\begin{equation}
B_\nu=\sum_{\bf q} C_\nu({\bf q})\,b_{\bf q}
 \label{B}
\end{equation}
with
\begin{equation}
  C_\nu({\bf q})=\frac{1}{\sqrt{\Lambda_\nu}}\sum_n F_n({\bf q})X_{\nu n}
\end{equation}
making the transformation orthogonal,
$$\sum_{\bf q}
C^\ast_\nu({\bf q}) C_\mu({\bf q})=\delta_{\nu\mu}\,.
$$
Our calculation of the optical polarization $P(t)$ also needs to
adapt the unshifted operators $a_{\bf q}$ to the form of
Eq.(\ref{B}) by making a similar transformation,
\begin{equation}
A_\nu = \sum_{\bf q} C_\nu({\bf q}) a_{\bf q}\,,
\end{equation}
 which preserves
diagonality of the unperturbed Hamiltonian due to negligible LO
phonon dispersion:
\begin{equation}
H_0 = \omega_0\sum_\nu \left(A^\dagger_\nu
A_\nu+\frac{1}{2}\right).
\end{equation}
 For the
perturbed Hamiltonian $H_1$, the diagonalization procedure is
finalized as follows
\begin{eqnarray}
H_1-\widetilde{E}_1&=&\sum_\nu\left[\omega_0 \left(B^\dagger_\nu
B_\nu+\frac{1}{2}\right)-\frac{1}{2}\Lambda_\nu(B_\nu+B^\dagger_\nu)^2\right]
\nonumber\\
&=& \sum_\nu\Omega_\nu \left(\widetilde{A}^\dagger_\nu
\widetilde{A}_\nu+\frac{1}{2}\right)\,,
 \label{H1step3}
\end{eqnarray}
where we have switched to another Bose operators
\begin{equation}\widetilde{A}_\nu=\alpha^+_\nu B_\nu+\alpha^-_\nu
B^\dagger_\nu=\alpha_\nu^+ A_\nu+\alpha^-_\nu
A^\dagger_\nu+\psi_\nu
 \end{equation} with
\begin{eqnarray}
&&\Omega_\nu=\sqrt{\omega_0^2-2\omega_0\Lambda_\nu}\,,\label{Om}\\
&&
 \alpha^\pm_\nu=\frac{\Omega_\nu\pm\omega_0}{\sqrt{\Omega_\nu\omega_0}}\,,
\nonumber
\\
&& \psi_\nu=\alpha^+_\nu \phi_\nu+\alpha^-_\nu \phi^\ast_\nu\,,\nonumber\\
&& \phi_\nu=\sum_{\bf q} C_\nu({\bf q}) f_{\bf
 q}=\frac{1}{\omega_0-2\Lambda_\nu} \sum_{\bf q} C_\nu({\bf q}) M_{11}({\bf
 -q})\,. \nonumber
\end{eqnarray}

Now the linear polarization obviously becomes a product of
contributions from independent phonon subsystems (labelled by the
index $\nu$):
\begin{eqnarray}
P(t)&=&\theta(t) e^{-i\widetilde{E}_1 t}\nonumber  \\
&\times&\prod_\nu \sum_{kl} \gamma\exp\{-\beta\epsilon_k^\nu
-i(\varepsilon_l^\nu-\epsilon_k^\nu)t\}\, |\langle
k;\nu|l;\nu\!\!>\!\!|^2,\nonumber\\
\label{P}
\end{eqnarray}
where $\varepsilon_l^\nu=(l+1/2)\Omega_\nu$ and
$\epsilon_k^\nu=(k+1/2)\omega_0$ are,
respectively, eigenenergies of the partial Hamiltonians
$H_{1\nu}$ and $H_{0\nu}$:
\begin{eqnarray}
H_1-\widetilde{E}_1&=&\sum_\nu H_{1\nu}\,,\ \ \ \ \ \
  H_{1\nu} |l;\nu\!\!> =\varepsilon_l^\nu |l;\nu\!\!> ,
\\
H_0&=&\sum_\nu H_{0\nu}\,,\ \ \ \ \ \
 H_{0\nu}|k;\nu\rangle=\epsilon_k^\nu |k;\nu\rangle,
\end{eqnarray}
$\beta=1/k_BT$, and $\gamma=2\cosh(\beta\omega_0/2)$. The
projections $z^l_k(\nu)=\langle k;\nu|l;\nu\!\!>$ of the perturbed
phonon states ($\widetilde{A}_\nu$) into the bare ones ($A_\nu$)
are calculated recursively (below the index $\nu$ is omitted for
simplicity; $k,l\geq0$):
\begin{eqnarray}
 \sqrt{l+1}\, z_k^{l+1}&=&\alpha^+ \sqrt{k}\, z_{k-1}^l +\alpha^-
\sqrt{k+1}\, z_{k+1}^l+\psi^\ast z_k^l\,,
\nonumber\\
\sqrt{l} \,z_k^{l-1}&=&\alpha^+ \sqrt{k+1} \,z_{k+1}^l +\alpha^-
\sqrt{k} \,z_{k-1}^l+\psi z_k^l\,, \nonumber
\end{eqnarray}
where the second equation can be used to determine the start
values for the recursion ($l=0$), together with a normalization
condition $\sum_{k=0}^\infty |z_k^0|^2=1$.

In case of two levels (only one excited state $n=2$ is taken into
account), the matrix $A_{nm}$ becomes a scalar [see Eq.(\ref{A})]
and thus only one phonon mode is renormalized. This phonon mode is
bound to the QD and has a new frequency $\Omega$ given by
Eq.(\ref{Om}). Due to the LO phonon degeneracy, all the other
phonon modes are decoupled and thus do not contribute to the
spectrum. In fact, the remaining phonon modes can be always
orthogonalized (with respect to the bound modes) without change of
their frequency. According to Eq.(\ref{P}), the spectrum
represents a set of discrete lines with a fine structure splitting
given by the difference between $\omega_0$ and $\Omega$, see
Fig.1\,(a). Increasing the number of exciton levels participating
in the coupling, the number of bound phonon modes (having
renormalized frequencies, Eq.(\ref{Om}), different from
$\omega_0$) also increases and thus delta lines in the spectrum
become more dense, see Fig.1\,(b) and (c). Finally, if we include
all exciton levels, these lines merge and produce a continuous
spectral broadening, Fig.1\,(d).

\begin{figure}[t]
\hspace{0.1cm}
\includegraphics*[angle=0,width=0.95\linewidth]{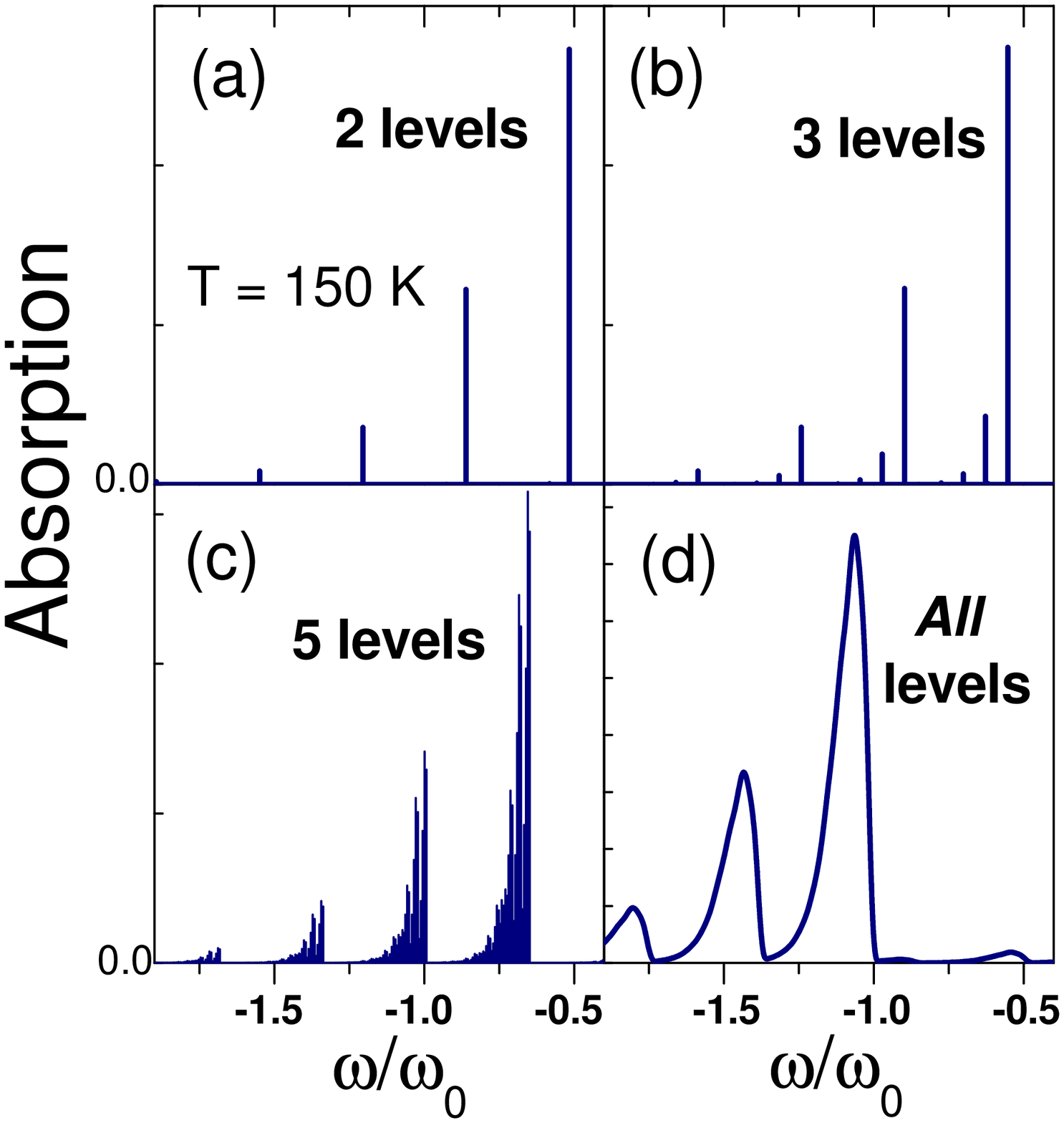}
  \caption{Calculated absorption spectra of the ground exciton state
in a CdSe QD in the model of infinite harmonic potentials with the
distance of 80\,meV (35\,meV) between electron (hole) levels
coupled to LO phonons \mbox{($\omega_0=24$\,meV).}}
\end{figure}

\begin{figure*}[h]
\hspace{2cm}\includegraphics*[angle=0,width=0.75\textwidth]{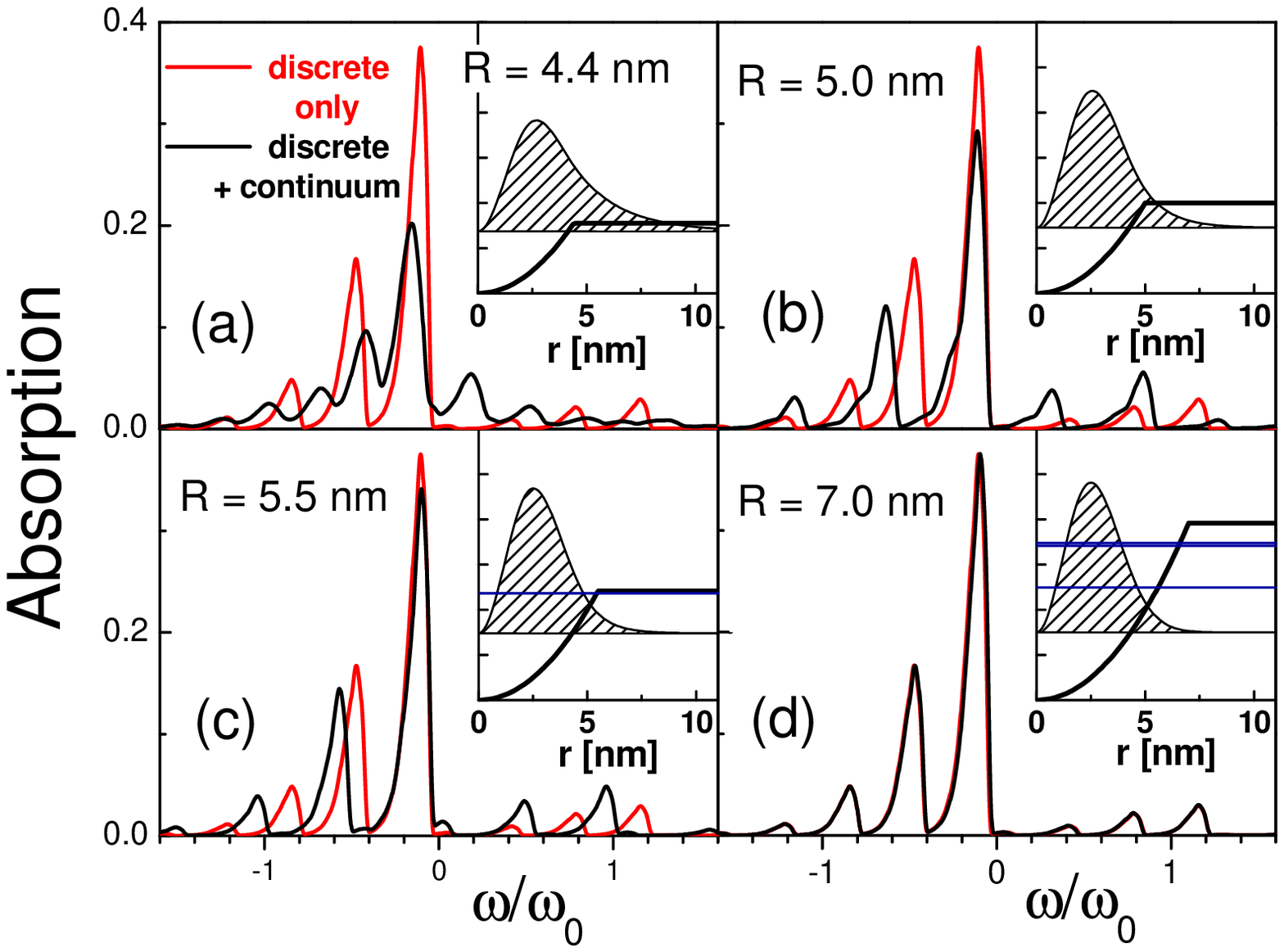}
\caption{Black curves: absorption calculated for different shapes
of QD potentials shown in the insets along with the ground state
wave function (shaded area) and discrete levels (horizontal lines)
for the electron. Red curves: the spectrum calculated in the model
of unlimited parabolic potentials, i.e. the same as in Fig.1\,(d).
Zero of energy is the polaron shifted ground state transition
energy; $T=150$\,K. }
\end{figure*}

This nontrivial mechanism of dephasing can be understood in terms
of bound phonon states~\cite{Dean70}. As shown in our recent
paper~\cite{Muljarov07}, the bound-phonon frequencies $\Omega_\nu$
are condensed just below $\omega_0$. Such an accumulation of
infinitely many levels becomes a mathematical clue for the
continuous spectral broadening. There are even some similarities
to the model of quadratic coupling to acoustic pho\-nons, in which
taking into account only one excited state already results in
infinitely many bound-pho\-non modes with re\-nor\-malized
frequencies and finally in a spectral
broadening~\cite{Muljarov04}. However, the flat dispersion of LO
phonons makes the polarization decay essentially non-exponential,
in contrast with purely exponential long-time behavior caused by
acoustic phonons~\cite{Muljarov04,Borri01}.

The absorption spectra in Fig.1 are calculated for a QD with CdSe
parameters, in the model of spherical harmonic potentials having
only discrete levels. Now we analyze how the absorption of the QD
changes if, instead of the unlimited parabolic confinement, a more
realistic potential (parabola truncated at $r=R$) is used, so that
the electron (and the hole) has both discrete and continuum
states, see insets in Fig.2. Below $R=5.5$\,nm [Fig.2\,(a,b)] only
one discrete level exists, while for $R=5.5$\,nm and 7\,nm
[Fig.2\,(c) and (d)] there are, respectively, one and three
excited discrete states (shown by green horizontal lines).
All other excited states (not shown) belong to the wetting layer
(WL) continuum. States in the continuum are modelled in a large
sphere with rigid walls. Numerically, it is checked that for the
given accuracy, there is no dependence on the number of continuum
states included in the model and on the radius of the large
sphere. In the worst case we have to take into account about 1500
densely spaced states in a 100\,nm sphere. Below $R=7$\,nm the
spectra for truncated parabolas (black curved) are obviously
different from that of the unlimited parabola (red curves), see
Fig.2\,(a) to (c). However, the linewidth remains practically the
same. Finally, for $R\geq7$\,nm, the two spectra almost coincide,
Fig.2\,(d).

In conclusion, a novel mechanism of spectral broadening in QDs due
to the standard linear exciton-LO phonon coupling is found and
demonstrated by mapping it into exactly solvable quadratic
coupling model. Generalizing this model to an arbitrary number of
exciton excited states, the resulting Hamiltonian is diagonalized
exactly and the optical polarization and absorption are calculated
in terms of QD-bound phonon modes. It is shown that only taking
into account the full excitonic spectrum, i.e. the complete
manifold of excited states including both confined and the
WL-continuum states, results in a LO phonon-induced exciton
dephasing and spectral broadening. In simulation of the absorption
in deep QDs with a few confined exciton states, the WL continuum
can be safely replaced by a discrete spectrum, while in shallow
QDs the continuum states have to be properly taken into account.

\begin{acknowledgement}
E.\,A.\,M. acknowledges partial support by Russian Foundation for
Basic Research and Russian Academy of Sciences.
\end{acknowledgement}

\end{document}